\documentclass[11pt]{article}
\usepackage{latexsym}
\usepackage{times}
\usepackage{enumerate}
\usepackage{mathrsfs}
\usepackage{stmaryrd}
\usepackage{amsopn}
\usepackage{amsmath}
\usepackage{amssymb}
\usepackage{amsfonts}
\usepackage{amsbsy}
\usepackage{amscd}

at 11pt \font\sevenrsf=rsfs7 at 8pt \font\fiversf=rsfs5 at 6pt
\newfam\rsffam
\textfont\rsffam=\tenrsf \scriptfont\rsffam=\sevenrsf
\scriptscriptfont\rsffam=\fiversf

\newtheorem{theo}{\rm \bf Theorem}[section]
\newtheorem{lemme}[theo]{Lemma}
\newtheorem{propo}[theo]{Proposition}

\newtheorem{defi}[theo]{Definition}

\newtheorem{nb}[theo]{Remark}

\font\teneuf=eufm10 at 12pt \font\seveneuf=eufm7 at 8pt
\font\fiveeuf=eufm5 at 6pt
\newfam\euffam
\textfont\euffam=\teneuf \scriptfont\euffam=\seveneuf
\scriptscriptfont\euffam=\fiveeuf

\newfont{\secgoth}{eufm10 at 16pt}
\newenvironment{preuve}{\noindent {\tt Proof:}}{\hfill$\blacksquare$\bigskip}
\def \bq {\begin{equation}}
\def \eq {\end{equation}}

\def \leq {\leqslant}
\def \geq {\geqslant}

\newenvironment{proof1}{\noindent {\tt Proof of Theorem \ref{main}:}}{\hfill$\blacksquare$\bigskip }
\newenvironment{proof2}{\noindent {\tt Proof of Theorem \ref{main2}:}}{\hfill$\blacksquare$\bigskip}

\def \v {v_{\star}}
\def \d {\mathrm{d}}
\def \H {\mathbf{H}}
\def \sqx {\sqrt{x^2+y^2}}
\def \sqz {\sqrt{z^2+\y^2}}
\def \sqr {\sqrt{R^2-y^2}}
\def \sqry {\sqrt{R^2-\y^2}}

\def \D {\mathrm{Dom}}

\def \y {\eta}
\def \ut {(U(t))_{t \geq 0}}
\def \uqt {(U_q(t))_{t \geq 0}}

\def \vt {(V(t))_{t \geq 0}}

\def \xs {\xi_{\star}}

\renewcommand{\epsilon}{\varepsilon}

\numberwithin{equation}{section}

\begin{document}

\title{\sf Stability of the essential spectrum for $2D$--transport models with Maxwell boundary
conditions.
 }

\author{{\bf Bertrand Lods} \\ \normalsize Politecnico di Torino,
Dipartimento di Matematica,\\ \normalsize Corso Duca degli Abruzzi,
24,
10129 Torino, Italia. \\
\normalsize E-mail: {\tt lods@calvino.polito.it}\\
\\
{\bf Mohammed Sbihi} \\
\normalsize Universit\'e de Franche-Comt\'e, Equipe de
Math\'ematiques CNRS UMR 6623, \\ \normalsize 16, route
de Gray, 25030 Besan\c{c}on Cedex, France.\\
\normalsize E-mail: {\tt msbihi@math.univ-fcomte.fr} }

\date{}
\maketitle
\begin{abstract}We discuss the spectral properties of collisional semigroups
associated to various models from transport theory by exploiting
the links between the so-called resolvent approach and the
semigroup approach. Precisely, we show that the essential spectrum
of the full transport semigroup coincides with that of the
collisionless transport semigroup in any $L^p$--spaces $(1 <p <
\infty)$ for three $2D$--transport models with Maxwell--boundary
conditions.
\end{abstract}


\section{Introduction}
%
This work follows the very recent one of the first author on
several mono--energetic transport problems \cite{ttsp} by dealing
now with collisional models. Precisely, we show  that the
essential spectrum of the full transport semigroup coincides with
that of the collisionless transport semigroup associated to the
following $2D$--models:
\begin{enumerate}[i)\:]
\item The Rotenberg model with boundary conditions of Maxwell
type. \item The one--velocity transport equation in a sphere with
Maxwell--type boundary conditions. \item The mono--energetic
transport equation in a slab of thickness $2a > 0$.
\end{enumerate}
These three models are particular versions of the more general
transport equation
\begin{equation}\label{1a} \dfrac{\partial \phi}{\partial t}(x,\xi,t)+ \xi \cdot \nabla_x
\phi(x,\xi,t)+\sigma(x,\xi)\phi(x,\xi,t)=\int_V\kappa(x,\xi,\xi_{\star})\phi(x,\xi_{\star},t)
\d \nu(\xi_{\star}),\\
\end{equation}
with the initial condition
\begin{equation}\label{1b}\phi(x,\xi,0)=\phi_0(x,\xi) \qquad \qquad (x,\xi) \in \Omega
\times V \end{equation} and with Maxwell boundary conditions
\begin{equation}\label{1c}
\phi_{|\Gamma_-}(x,\xi,t)=H(\phi_{|\Gamma_+})(x,\xi,t) \qquad \qquad
(x,\xi) \in \Gamma_-, t >0
\end{equation}
where $\Omega$ is a smooth open subset of $\mathbb{R}^N$ $(N \geq
1)$, $V$ is the support of a positive Radon measure $\d\nu$ on
$\mathbb{R}^N$ and $\phi_0 \in X_p:=L^p(\Omega \times V,\d
x\d\nu(\xi))$ $(1 \leq p < \infty).$ Here $\Gamma_-$ (resp
$\Gamma_+$) denotes the incoming (resp. outgoing) part of the
boundary of the phase space $\Omega \times V$,
$\Gamma_{\pm}=\{(x,\xi) \;\in \partial \Omega \times V\;;\;\pm \xi
\cdot n(x) > 0\}$ where $n(x)$ stands for the outward normal unit
at $x \in \partial \Omega$. The boundary condition \eqref{1c}
expresses that the incoming flux $\phi_{|\Gamma_-}(\cdot,\cdot,t)$
is related to the outgoing one $\phi_{|\Gamma_+}(\cdot,\cdot,t)$
through a \textit{linear operator} $H$ that we shall assume to be
bounded on
some suitable trace spaces. 
The collision operator $\mathcal{K}$ arising at the
right-hand--side of \eqref{1a} is assumed to be a bounded operator
in $L^p(\Omega \times V,\d x\d \nu(\xi))$ $(1 \leq p < \infty)$
and it is well--known that $\mathcal{K}$ induces some compactness
with respect to the velocity $\xi$. The well--posedness of the
free--streaming version of \eqref{1a} (corresponding to null
collision $\kappa=0$) has been investigated recently in
\cite{CRAS,gener} where sufficient conditions on the boundary
operator $H$ are given ensuring that the free streaming operator
generates a $c_0$--semigroup $\ut$ in $L^p(\Omega \times V,\d x \d
\nu(\xi))$. Then, since $\mathcal{K}$ acts as a {\it bounded
perturbation} of $\ut$, the model \eqref{1a}--\eqref{1c} is
governed by a $c_0$--semigroup $\vt$ in $L^p(\Omega \times V,\d x
\d \nu(\xi))$.

It is well-known that the asymptotic behavior (as $t \to \infty$)
of the solution $\phi(\cdot,\cdot,t)$ to \eqref{1a}--\eqref{1c} is
strongly related to the spectral properties of the semigroup
$\vt$. In particular, an important task is the stability of the
essential spectrum \cite{schechter}: does \bq
\label{ess}\sigma_{\mathrm{ess}}(V(t))=\sigma_{\mathrm{ess}}(U(t))\eq
for any $t \geq 0$ ?

This question has been answered positively in the case of
non--reentry boundary conditions (i.e. $H=0$) in
\cite{mmkprep1,sbihi} by showing that the difference $V(t)-U(t)$
is a compact operator in $L^p(\Omega \times V,\d x\d \nu(\xi))$
$(1 < p < \infty)$. The case of re--entry boundary conditions is
much more involved because of the difficulty to compute the
semigroup $\ut$ in this case. There exists a few partial results
dealing with the above models i)--iii) \cite{bouem1,
bouem2,chabi1,chabi2, green,jeribi,lalods,  zhang, zhang1} but the
asymptotic behavior of the solution to the associated equations is
investigated only for smooth initial data or, at best, estimates
of the essential type of the $\vt$ are provided. Therefore,
question \eqref{ess} is \textit{totally open} for the
aforementioned models.

The present paper generalizes the previous ones by establishing the
above identity \eqref{ess} for the three above models i)-iii) in the
case $1<p<\infty.$ The strategy is based upon the so--called
resolvent approach, already used in \cite{lehner, VImmklaplace,
lalods}, and exploits the link between this approach and the
compactness of $V(t)-U(t)$ recently discovered, in \cite{brendle,
huang, sbihi} (see Section 2 for more details). Note that the
results of \cite{brendle, sbihi} are valid only \textit{in a Hilbert
space setting} but we will see in this paper how they allow to treat
the above three above models in any $L^p$--space with $1 < p <
\infty.$
Indeed, for \textit{Maxwell-like boundary conditions} in
$2D$--geometry, the boundary operator $H$ splits as
$H=\mathcal{J}+K$ where $\mathcal{J}$ is a multiplication operator
and $K$ is compact. This allows to approximate it by some
finite--rank operators. Under some natural assumptions on the
collision operator $\mathcal{K}$, it is then possible to
approximate both $U(t)$ and $V(t)$ in such a way that both of them
are bounded operator in any $L^r(\Omega \times V,\d x\d
\nu(\xi))$, $1 < r < \infty$. Then, by an interpolation argument,
it is sufficient to prove the compactness of the difference
$V(t)-U(t)$ in the Hilbert space $L^2(\Omega \times V)$. This
strategy excludes naturally the case $p=1$ for which a specific
analysis is necessary.\smallskip

Let us explain more in details the content of the paper. In the
following section, we present the resolvent approach and the result
of the second author we shall use in the rest of the paper. In
section 3, we investigate the Rotenberg model and gave a precise
description of the method of the proof of identity \eqref{ess}. In
section 4, we deal with the mono--energetic transport equation in a
sphere by adopting the approach exposed in Section 3. Finally, we
deal in section 5 with the model iii).\medskip

\noindent {\bf Notations.} Given two Banach spaces $X$ and $Y$,
$\mathfrak{B}(X,Y)$ shall denote the set of bounded linear
operators from $X$ to $Y$ whereas the ideal of compact operators
from $X$ to $Y$ will be denoted $\mathfrak{C}(X,Y)$. When $X=Y$,
we will simply write $\mathfrak{B}(X)$ and $\mathfrak{C}(X)$.

\section{On the resolvent approach}

We recall here the link between the so--called resolvent approach
and the study of the compactness of the difference of semigroups.
Let $X$ be a Banach space and let $T$ : $\D(T) \subset X \to X$ be
the infinitesimal generator of a $c_0$-semigroup of operators
$\ut$ in $X$. We consider the Cauchy problem
\begin{equation}\label{cauchy}
\begin{cases}
\dfrac{\mathrm{d} \phi}{\mathrm{d} t}(t)&=\left(T+K\right)\phi(t) \qquad t \geqslant  0,\\
\phi(0)&=\phi_0
\end{cases}
\end{equation}
where $K \in \mathfrak{B}(X)$ and $\phi_0 \in X.$ Since $A:=T+K$
is a bounded perturbation of $T$, it is known that $A$ with domain
$\D(A)=\D(T)$ generates a $c_0$-semigroup $\vt$ on $X$ given by
the Dyson--Phillips expansion \bq \label{dyson}
V(t)=\sum_{j=0}^{\infty}U_j(t)\eq where $U_0(t)=U(t),$
$U_j(t)=\displaystyle  \int_0^t U(t-s)KU_{j-1}(s)ds$ $(j \geqslant
1)$.\medskip

\noindent When dealing with the time--asymptotic behavior of the
solution $\phi(t)$ to \eqref{cauchy}, until recently, two
techniques have been used. The first one, called the
\textit{semigroup approach}, consists in studying the remainder
$R_n(t)=\sum_{j \geq n}U_j(t)$ of the Dyson--Phillips expansion
\eqref{dyson} (see \cite{vidav}). Actually, if there is $n
\geqslant 0$ such that $R_n(t) \in \mathfrak{C}(X)$ for any $t
\geq 0$ then $\sigma(V(t)) \cap \{\mu \in \mathbb{C}\,;\,|\mu|
> \exp(\eta t)\}$ consists of, at most, isolated eigenvalues with
finite algebraic multiplicities where $\eta$ is the type of $\ut$.
Therefore, for any $\nu
> \eta$, $\sigma(T) \cap \{\mathrm{Re} \lambda > \nu\}$ consists
of a finite set of isolated eigenvalues
$\{\lambda_1,\ldots,\lambda_n\}.$ Then, the solution to
\eqref{cauchy} satisfies
\begin{equation}\label{beta}
\lim_{t \to \infty} \exp(-\beta t)\|\phi(t)-\sum_{j=1}^n \exp(\lambda_j t + D_j
t)P_j\phi_0\|=0
\end{equation}
where $\phi_0 \in X,$ $P_j$ and $D_j$ denote, respectively, the
spectral projection and the nilpotent operator associated to
$\{\lambda_i,\,i=1,\ldots,n\}$ and
$$\sup\{\mathrm{Re}\lambda, \,\lambda \in \sigma(T_H), \,\mathrm{Re}\lambda < \nu\} <
\beta < \min\{\mathrm{Re}\lambda_j,\,j=1,\ldots,n\}.$$ Of course,
the success of such a method is strongly related to the
possibility of computing the terms of the Dyson--Phillips
expansion \eqref{dyson}. Until recently, it appeared to be the
only way to discuss their compactness properties. Unfortunately,
in practical situations, the unperturbed semigroup $\ut$  may not
be explicit or at least can turn out to be hard to handle.

An alternative way to determine the long--time behavior of
$\phi(t)$ is the so--called \textit{resolvent approach} initiated
by J. Lehner and M. Wing \cite{lehner} in the context of neutron
transport theory and consists in expressing $\phi(t)$ as an
inverse Laplace transform of $(\lambda-T-K)^{-1}\phi_0$. This
method has been developed subsequently in an abstract setting by
M. Mokhtar--Kharroubi \cite{VImmklaplace} and, more recently by
Degong Song \cite{VIds} (see \cite{lalods} for an application of
the results of \cite{VIds} in the context of neutron transport
equation on a slab). The main drawback of this approach is that
\eqref{beta} is valid only for smooth initial data $\phi_0 \in
D(A)$. In particular, even in Hilbert spaces, it does not permit
to explicit the essential type of $V(t)$ but only to give some
estimates of it \cite{VIds}.

In a Hilbert space setting, these two approaches have been linked
recently by S. Brendle  \cite{brendle}. Precisely, if there exist
some $\alpha > w_0(U)$ and some integer $m$ such that

$$(\lambda-T)^{-1}\left(K(\lambda-T)^{-1}\right)^m \quad \text{ is
compact for any Re}\lambda=\alpha$$ and
$$\lim_{|\textrm{Im}\lambda| \to
\infty}\|(\lambda-T)^{-1}\left(K(\lambda-T)^{-1}\right)^m\|=0
\qquad \forall \textrm{Re}\lambda=\alpha$$ then the
$(m+2)$--remainder term $R_{m+2}(t)$ of the Dyson--Phillips
expansion series is compact. Such a result, though really
important for the applications, does not allow to investigate the
compactness of the difference of the two semigroups
$V(t)-U(t)=R_1(t).$ Very recently, the second author, inspired by
the work of S. Brendle \cite{brendle}, has been able to provide
sufficient conditions in terms of the resolvent of $T$ ensuring
the compactness of the \textit{first remainder term} $R_1(t)$.
Precisely \cite[Corollary 2.2, Lemma 2.3]{sbihi},
\begin{theo}\label{sbihi}
Assume that $T$ is dissipative and there exists $\alpha
> w_0(U)$ such that \bq \label{comp}
(\alpha+i\beta-T)^{-1}K(\alpha+i\beta-T)^{-1} \quad \text{ is
compact for all } \beta \in \mathbb{R} \eq and \bq \label{imlam}
\lim_{|\beta| \to \infty}
\left(\|K^{\star}(\alpha+i\beta-T)^{-1}K\|+\|K(\alpha+i\beta-T)^{-1}K^{\star}\|\right)=0\eq
then $R_1(t)$ is compact for all $t \geq 0$. In particular,
$\sigma_{\mathrm{ess}}(V(t))=\sigma_{\mathrm{ess}}(U(t)).$
\end{theo}

We refer to \cite{sbihi} for a proof of this result as well as for
its application to neutron transport equation in bounded geometry
with absorbing boundary conditions.
\begin{nb}\label{crit} Actually, under the
hypothesis \eqref{imlam}, the mapping $t \geq 0 \mapsto R_1(t) \in
\mathfrak{B}(X)$ is continuous \cite{sbihi}. This implies the
stability of the critical spectrum (see \cite{poland} for a
precise definition)
$\sigma_{\mathrm{crit}}(V(t))=\sigma_{\mathrm{crit}}(U(t))$ for
any $t \geq 0$. Such an identity plays a crucial role for
establishing spectral mapping theorems (see \cite{mmksbihi} for a
recent application to neutron transport equations in unbounded
geometries).\end{nb}

\begin{nb}\label{stationn} Note that Assumption \eqref{comp} implies that $(\lambda-T - K)^{-1}-(\lambda-T)^{-1} \in \mathfrak{C}(X)$ for
any $\lambda \in \rho(T+K)$. Therefore,
$\sigma_{\mathrm{ess}}(T+K)=\sigma_{\mathrm{ess}}(T)$.\end{nb}

Let us recall now the definition of regular collision operators as
they appear in \cite{mmkbook}. The notations are those of the
introduction.
\begin{defi}\label{regular}
An operator $\mathcal{K} \in \mathfrak{B}(X_p)$ $(1 < p < \infty)$
is said to be regular if $\mathcal{K}$ can be approximated in the
norm operator by operators of the form:
\begin{equation}\label{finiterank}
\varphi \in X_p \mapsto \sum_{i \in
I}\alpha_i(x)\beta_i(\xi)\int_V
\theta_i(\xi_{\star})\varphi(x,\xi_{\star})\d \nu(\xi_{\star}) \in
X_p
\end{equation}
where $I$ is finite, $\alpha_i \in L^{\infty}(\Omega)$, $\beta_i
\in L^p(V,\d \nu(\xi))$ and $\theta_i \in L^q(V,\d \nu(\xi))$,
$1/p+1/q=1.$\end{defi}
\begin{nb}\label{nbregular} Since $1 < p < \infty$, one notes that the set $\mathcal{C}_c(V)$ of continuous functions with
compact support in $V$ is dense in $L^q(V,\d \nu(\xi))$ as well as
in $L^p(V,\d \nu(\xi))$ ($1/p+1/q=1$). Consequently, one may
assume in the above definition that $\beta_i(\cdot)$ and
$\theta_i(\cdot)$ are continuous functions with compact supports
in $V$.\end{nb}

We end this section with a simple generalization of the classical
Riemann--Lebesgue Lemma we shall invoke often in the sequel.

\begin{lemme}\label{riemann}
Let $f \in L^1(\mathbb{R}) \cap L^{\infty}(\mathbb{R})$ be
compactly supported on some interval $]a,b[ \subset \mathbb{R}$ $(
a < b < \infty)$. Let $\omega(\cdot)$ be a bijective and
continuously differentiable function on $\mathbb{R}$ whose
derivative admits a finite number of zeros on $]a,b[.$ Then,
$$\lim_{|\xi| \to \infty}\int_{\mathbb{R}} e^{i\xi \omega(x)}
\,f(x)\,\d x=0.$$
\end{lemme}

\begin{preuve} Let us denote by $\omega'(\cdot)$ the derivative of
$\omega(\cdot)$ and assume, without loss of generality, that there
is a unique $x_0 \in \mathbb{R}$ such that $\omega'(x_0)=0.$ Let
$\epsilon > 0$ be fixed. Since $f \in L^1(\mathbb{R}) \cap
L^{\infty}(\mathbb{R})$, there exists
$\delta=\epsilon/2\|f\|_{\infty}
> 0$ such that $$\sup_{\xi \in \mathbb{R}}
\left|\int_{x_0-\delta}^{x_0+\delta} e^{i \xi \omega(x)} f(x) \d
x\right| \leq \epsilon.$$ Consequently, it is enough to prove
that, for sufficiently large $|\xi|$, \bq \label{rie}
\left|\int_{x \leq x_0-\delta}e^{i\xi \omega(x)} \,f(x)\,\d
x\right| + \left|\int_{x \geq x_0 + \delta}e^{i\xi \omega(x)}
\,f(x)\,\d x\right| \leq 2\epsilon.\eq Let us deal with the first
integral. Since $\omega(\cdot)$ is bijective and $f$ is compactly
supported in $]a,b[$,
$$\int_{x \leq
x_0-\delta}e^{i\xi \omega(x)} \,f(x)\,\d
x=\int_{x_0-\delta}^{b}e^{i\xi \omega(x)} \,f(x)\,\d x=\int_I e^{i
\xi y} f(\omega^{-1}(y))\dfrac{\d y}{\omega'(\omega^{-1}(y))}$$
where $I=\omega^{-1}([x_0-\delta,b])$ is a compact interval. Now,
since $\omega'\neq 0$ on $[x_0-\delta,b]$, it is clear that
$$G(\cdot):=f(\omega^{-1}(\cdot))\dfrac{1}{\omega'(\omega^{-1}(\cdot))} \in
L^1(I)$$ and the (classical) Riemann--Lebesgue Lemma asserts that
$$\lim_{|\xi| \to \infty}\int_I e^{i \xi y}G(y)\d y=0.$$
One proceeds in the same way with the second integral of
\eqref{rie} and this ends the proof.\end{preuve}

\section{On the Rotenberg model}

\subsection{Statement of the result}

We first consider a model of growing cell populations proposed by
Rotenberg in 1983 \cite{roten} as an improvement of the
Lebowitz--Rubinow model \cite{Lebowitz}. Each cell is characterized
by its \textit{degree of maturity} $\mu$ and its maturation velocity
$v=\frac{d\mu}{dt}.$ The degree of maturation is defined so that
$\mu=0$  at birth (\textit{daughter-cells}) and $\mu=1$ when the
cell divides by mitosis (\textit{mother-cells}). The second variable
$v$ is considered as an independent variable within $]a,b[$ $(0 \leq
a < b \leq \infty)$. Denote by $f(t,\mu,v)$ the density of cells
having the degree of maturity $\mu$ and maturation velocity $v$ at
time $t \geq 0$. It satisfies the following transport-like equation:
\begin{multline}\label{rot}
\dfrac{\partial f}{\partial t}(t,\mu,v)+v\dfrac{\partial f}{\partial
\mu}(t,\mu,v)+\sigma(\mu,v)f(t,\mu,v)=\int_a^br(\mu,v,v')f(t,\mu,v')dv'\\
\qquad (\mu,v) \in ]0,1[ \times ]a,b[, t \geq 0;
\end{multline}
where the kernel $r(\mu,v,v')$ is the transition rate at which cells
change their velocities from $v'$ to $v$ and $\sigma(\mu,v)$ denotes
the \textit{mortality rate}. During the mitosis, three different
situations may occur. First, one can assume that there is a positive
correlation $k(v,v') \geq 0$ between the maturation velocity $v'$ of
a "mother-cell" and the one $v$ of a "daughter-cell". In this case
the reproduction rule is given by
\begin{equation}\label{diff}
vf(t,0,v)=\alpha\int_{a}^{b}k(v,v')f(t,1,v')v'dv' \qquad v \in
(a,b),
\end{equation}
where $\alpha \geq 0$ is the average number of viable daughters
per mitosis. Second, one can assume that daughter cells perfectly
inherit their maturation velocity from mother (\textit{perfect
memory}), i.e. $v=v'$, or equivalently $k(v,v')=\delta(v-v')$
where $\delta(\cdot)$ denotes the Dirac mass at zero. Then, the
biological reproduction rule reads:
\begin{equation*}
f(t,0,v)=\beta(v)\;f(t,1,v) \qquad v \in ]a,b[,
\end{equation*}
where $\beta(v) \geq 0$ denotes the average number of viable
daughters per mitosis. Finally, one can combine the two previous
transition rules which leads to the general reproduction rule we
will investigate in the sequel:
\begin{equation}\label{rotbc}
f(t,0,v)=\beta(v)\,f(t,1,v)+\dfrac{\alpha}{v}\int_{a}^{b}k(v,v')f(t,1,v')v'dv'
\qquad v \in ]a,b[.
\end{equation}
Of course, one has to complement \eqref{rot} and \eqref{rotbc} with
an initial condition
\begin{equation}\label{rotini}
f(0,\mu,v)=f_0(\mu,v) \qquad v \in ]a,b[,\;\mu \in ]0,1[
\end{equation}
where $f_0 \in X_p=L^p(]0,1[ \times ]a,b[;d\mu dv)$ $(1 < p <
\infty)$. The above model has been numerically solved by Rotenberg
\cite{roten}. The first theoretical approach of this model can be
found in the monograph \cite[Chapter XIII]{proto}. Later, this
model has been investigated in \cite{bouem1,bouem2}. The
asymptotic behavior of the solution to \eqref{rot}--\eqref{rotini}
has been dealt with in \cite{jeribi} for diffuse boundary
conditions \eqref{diff} and for a smooth initial data. We will
generalize the result of \cite{jeribi} by dealing with the more
general reproduction rule \eqref{rotbc} and by showing the
stability of the essential spectrum. Let us make the following
assumptions:

$(\mathrm{H}1)$ The collision operator
$$B\::\phi \mapsto B\phi(\mu,v)=\int_a^br(\mu,v,v')\phi(\mu,v')dv'$$
is a bounded and nonnegative operator in $X_p$ $(1 < p < \infty)$.

$(\mathrm{H}2)$ The mortality rate $\sigma(\cdot,\cdot)$ is
bounded and nonnegative on $]0,1[ \times ]a,b[.$ We denote by
$\underline{\sigma}=\inf\{\sigma(\mu,v)\,;\,\mu \in ]0,1[, v \in
]a,b[\}.$

$(\mathrm{H}3)$ The kernel $k(\cdot,\cdot)$ is nonnegative and
such that the mapping $$K\;:\;f \in Y_p \mapsto \dfrac{\alpha}{v}
\int_a^bk(v,v')f(v')v'dv' \in Y_p$$ is compact, where
$Y_p=L^p(]a,b[,v\d v)$ $(1 < p < \infty).$

$(\mathrm{H}4)$ $0 \leq \beta(v) \leq \beta_0 < 1$ and $\alpha
\geq 0.$\medskip

\noindent Let us define the boundary operator $H \in
\mathfrak{B}(Y_p)$ by \bq \label{defH}Hf(v)=\beta(v)
f(v)+\dfrac{\alpha}{v}\int_{a}^{b}k(v,v')f(v')v'dv' =\beta(v)
f(v)+ Kf(v) \qquad f \in Y_p.\eq Define the unbounded operator
$\textbf{A}_{H}$ by
$$\textbf{A}_{H}\phi(\mu,v)=-v \frac{\partial
\phi}{\partial \mu}(\mu,v)-\sigma(\mu,v)\phi(\mu,v)$$ with domain
$\D(\textbf{A}_{H})$ given by
$$
\{\phi \in X_p \text{ such that } \textbf{A}_{H}\phi \in X_p,
\phi(0,v) \text{ and } \phi(1,v) \in Y_p \text{ and satisfy }
\eqref{rotbc}\}.$$ Note that, since $K \in \mathfrak{C}(Y_p)$ and
$\beta_0 < 1$, \cite[Theorem 6.8]{gener} implies the following:
\begin{theo} Assume $(\mathrm{H}1)-(\mathrm{H}4)$ to be fulfilled.
Then, $\mathbf{A}_H$ generates a nonnegative $c_0$--semigroup
$\ut$ of $X_p$ $(1 < p < \infty)$. As a consequence,
$\textbf{A}_H+B$ is also the generator of a $c_0$--semigroup $\vt$
of $X_p$.\end{theo}
\begin{nb} Note that a complete description of the spectrum of
$\mathbf{A}_H$ is provided in \cite{ttsp}.\end{nb} Concerning the
asymptotic behavior of $\vt$, one states the following:
\begin{theo}\label{main} Let $1 < p < \infty$ and let $B \in \mathfrak{B}(X_p)$
be regular. Then, $V(t)-U(t)$ is compact for any $t \geq 0$ and
$\sigma_{\mathrm{ess}}(U(t))=\sigma_{\mathrm{ess}}(V(t))$ for any
$t \geq 0.$ In particular, $\sigma_{\mathrm{ess}}(\mathbf{A}_H +
B)=\sigma_{\mathrm{ess}}(\mathbf{A}_H)=\sigma(\mathbf{A}_H)$.
\end{theo}
\begin{nb} We point out that Theorem \ref{main} covers all the
possible choice of the parameters $a,$ and $b$, namely $0 \leq a
\leq b \leq \infty$.\end{nb}

%

\subsection{Proof of Theorem \ref{main}}\label{secmain}

All this section is devoted to the proof of Theorem \ref{main}. As
a first step, one sees that the mortality rate does not play any
role in the compactness of the remainder $R_1(t)$. Indeed, let
$\widetilde{\mathbf{A}_H}$ stands for the operator $\mathbf{A}_H$
associated to the constant mortality rate $\underline{\sigma}$.
Since $\widetilde{\mathbf{A}_H}-A_H$ is the multiplication
operator by the nonnegative function
$\sigma(\cdot,\cdot)-\underline{\sigma}$, the Dyson--Phillips
formula \eqref{dyson} implies that 
$U_H(t) \leq \widetilde{U_H}(t)$ for any $t \geq 0$ where
$(\widetilde{U_H}(t))_{t \geq 0}$ is the $c_0$--semigroup
generated by $\widetilde{\mathbf{A}_H}$. The same occurs for the
semigroup $(\widetilde{V_H}(t))_{t \geq 0}$ generated by
$\widetilde{\mathbf{A}_H}+\mathcal{K}$. Consequently, the first
remainder terms $R_1(t)$ and $\widetilde{R_1(t)}$ are such that
$$R_1(t) \leq  \widetilde{R_1(t)} \qquad \forall \,t \geq 0.$$
By a domination argument \cite{IVdo}, the compactness of
$\widetilde{R_1(t)}$ implies that of $R_1(t)$. Therefore, in order
to apply Theorem \ref{sbihi}, one may assume without loss of
generality that
$$\sigma(\mu,v)=-\underline{\sigma} \qquad \forall \:(\mu,v) \in
]0,1[ \times ]a,b[.$$

Now, we point out that it suffices to prove Theorem \ref{main} for
contractive boundary operator $\|H\|< 1.$ Indeed, if $\|H\| \geq
1$, recall  \cite{gener} that the semigroup $\ut$ enjoys the
following similarity property: There exists $q \in (0,1)$ such
that
$$U(t)=M_q^{-1}U_q(t)M_q\,\qquad t\geq 0 $$
where $M_q \in \mathfrak{B}(X_p)$ is invertible (see \cite{gener}
for details) and $\uqt$ is the $c_0$--semigroup generated by:
\begin{equation*}
\begin{cases}
\textbf{A}_{H_q}\::\: &\D(\textbf{A}_{H_q}) \subset X_p \to X_p\\
&\varphi \mapsto \textbf{A}_{H_q}\varphi(\mu,v)=-v \frac{\partial
\varphi}{\partial \mu}(\mu,v)-(\underline{\sigma} + \ln
q)\varphi(\mu,v)
\end{cases}
\end{equation*}
(note that the collision frequency associated to
$\textbf{A}_{H_q}$ is constant) where the boundary operator $H_q$
is given by $H_q\varphi(v)=H(\exp\{\ln q /v\}\varphi)(v)$. In
particular, $q \in (0,1)$ is such that $\|H_q\| < 1.$ With obvious
notations, one has
$$R_1(t)=M_q^{-1}R_{1,q}(t)M_q$$
and it suffices to prove the compactness of $R_{1,q}(t)$. From
now, we will assume that
$$\|H\| < 1.$$
Let us now explicit the resolvent of $\mathbf{A}_H$. To this aim,
for any Re$\lambda
> -\underline{\sigma}$, define
\begin{equation*}\begin{cases}
M_{\lambda}:&\: Y_p \longrightarrow Y_p\\
&u \longmapsto M_{\lambda}u(v)=u(v)\exp\{-\displaystyle
\frac{\lambda+\underline{\sigma}}{v}\},\end{cases}
\end{equation*}
\begin{equation*}\begin{cases}
\Xi_{\lambda}:&\: Y_p \longrightarrow X_p\\
&u \longmapsto \Xi_{\lambda}u(\mu,v)=u(v)\exp\{-\displaystyle
\frac{\mu}{v}(\lambda+\underline{\sigma})\},\end{cases}
\end{equation*}

\begin{equation*}\begin{cases}
G_{\lambda}:&\: X_p \longrightarrow Y_p\\
&\varphi \longmapsto
G_{\lambda}\varphi(v)=\displaystyle\frac{1}{v}\int_0^1\varphi(\mu',\v)\exp\{-\displaystyle
\frac{1-\mu'}{v}(\lambda+\underline{\sigma})\}\d \mu'\end{cases}
\end{equation*}
and

\begin{equation*}\begin{cases}
C_{\lambda}:&\: X_p \longrightarrow X_p\\
&\varphi \longmapsto C_{\lambda}\varphi(\mu,v)=\displaystyle
\frac{1}{v}\int_0^{\mu}\varphi(\mu',v)\exp\{-\dfrac{\mu-\mu'}{v}(\lambda+\underline{\sigma})\}\d
\mu'.\end{cases}
\end{equation*}

The resolvent of $\mathbf{A}_H$ is given by the following, whose
proof can be easily adapted from \cite{gener}.
\begin{propo}\label{reso}
Let $H \in \mathfrak{B}(Y_p)$ be given by \eqref{defH} where
$(\mathrm{H}1)-(\mathrm{H}4)$ are fulfilled. Then $\{\lambda \in
\mathbb{C}\,;\, Re\lambda
> - \underline{\sigma}\} \subset \rho(\mathbf{A}_H)$ and
\begin{equation}\label{eqres}
(\lambda-\mathbf{A}_H)^{-1}=\Xi_{\lambda}H(I-M_{\lambda}H)^{-1}G_{\lambda}+C_{\lambda}\qquad
Re\lambda > -\underline{\sigma}.\end{equation}
\end{propo}

An important fact to be noticed is that, though Theorem
\ref{sbihi} is a purely hilbertian result, it turns out to be
useful for the treatment of neutron transport problems in
$L^p$--spaces for any $1 < p < \infty$. The reason is the
following. Let $1 < p < \infty$ be fixed. We first note that
$R_1(t)$ depends continuously on the boundary operator $H \in
\mathfrak{B}(Y_p)$. Recalling that $H=\beta \mathrm{Id}+K$ where
$K$ is a compact operator on $Y_p$,
it suffices to prove the compactness of $R_1(t)$ for a
\textit{finite rank} operator $K$, i.e. we can assume without loss
of generality that the kernel $k(v,v')$ is a degenerate kernel of
the form: \bq \label{kdegen} k(v,v')=\sum_{j \in
J}g_j(v)k_j(v')\eq where $J \subset \mathbb{N}$ is finite,
$g_j(\cdot) \in L^p(]a,b[,v\d v)$ and $k_j(\cdot) \in
L^q(]a,b[,v\d v)$ $(1/p+1/q=1)$. Moreover, by density, one may
assume that $g_j(\cdot)$ and $k_j(\cdot)$ are continuous functions
with compact supports on $]a,b[$. In this case, one notes easily
that $H \in \mathfrak{B}(Y_r)$ for any $1 < r < \infty$ and the
same occurs for $(\lambda-\mathbf{A}_H)^{-1}$ according to
Proposition \ref{reso}. The Trotter--Kato Theorem implies then
that, for any $t \geq 0$, $U_H(t) \in
\bigcap_{1<r<\infty}\mathfrak{B}(X_r).$ Similarly, since $B$ is
regular and $R_1(t)$ depends continuously on $B \in
\mathfrak{B}(X_p)$, one may assume that $B$ is of the form
\eqref{finiterank} where, according to Remark \ref{nbregular}, the
functions $\beta_i$ and $\theta_i$ are continuous with compact
supports in $]a,b[$. In this case, it is easy to see that $B \in
\bigcap_{1 < r < \infty} \mathfrak{B}(X_r)$ so that the same
occurs for $R_1(t)$:
$$R_1(t) \in \bigcap_{1 < r < \infty} \mathfrak{B}(X_r).$$ Consequently, by an interpolation argument, if $R_1(t)$
is a compact operator on $X_2$, then $R_1(t)$ is compact on $X_p$
for any $1< p < \infty.$ With this procedure, we may restrict
ourselves to prove the compactness of $R_1(t)$ in $X_2$. In this
case, one has the following Proposition whose proof is postponed to
the Appendix of this paper.
\begin{propo}\label{iml} Let us assume that $p=2$. Then, for any regular
operator $B \in \mathfrak{B}(X_2)$ and any $\mathrm{Re}\lambda >
-\underline{\sigma}$:
$$\lim_{|\mathrm{Im}\lambda| \to \infty}
\left(\|B^{\star}(\lambda-\mathbf{A}_H)^{-1}B\|_{\mathfrak{B}(X_2)}
+\|B(\lambda-\mathbf{A}_H)^{-1}B^{\star}\|_{\mathfrak{B}(X_2)}\right)=0
.$$
\end{propo}
\medskip

\begin{proof1} We already saw that it suffices to prove the result for $p=2$.
Proposition \ref{iml} asserts that Property \ref{imlam} of Theorem
\ref{sbihi} is fulfilled. Moreover, according to \cite[Theorem
3.1]{latrach}, $B(\lambda-\mathbf{A}_H)^{-1}$ is compact for any
Re$\lambda > -\underline{\sigma}$. Since $\|H\| < 1$,
$\mathbf{A}_H$ is dissipative (see \cite{beals}), Theorem
\ref{sbihi} asserts that $R_1(t) \in \mathfrak{C}(X_2)$ and the
conclusion follows. The identity
$\sigma_{\mathrm{ess}}(\mathbf{A}_H+B)=\sigma_{\mathrm{ess}}(\mathbf{A}_H)=\sigma(\mathbf{A}_H)$
follows from Remark \ref{stationn} and \cite{ttsp}.
\end{proof1}

\section{On the mono-energetic transport equation in spherical
geometry}

\subsection{Statement of the result}

In this section we consider a one-velocity linear transport operator
with Maxwell--type boundary conditions in a spherical medium of
radius $R$. For this kind of geometry, neutron transport equation
reads \cite[Chapter 1]{agoshkov}:
\begin{equation*}\label{sph1}
\frac{\partial \phi}{\partial t}(r,\mu,t)+\mu \frac{\partial
\phi}{\partial r}(r,\mu,t)+\frac{1-\mu^2}{r}\frac{\partial
\phi}{\partial
\mu}(r,\mu,t)+\Sigma(r,\mu)\phi(r,\mu,t)=\mathcal{K}\phi(r,\mu,t)
\end{equation*}
with the boundary condition
\begin{equation}\label{sphbc} \phi(R,\mu,t)=\gamma(-\mu)
\phi(R,-\mu,t)+\int_0^1\kappa(\mu,\mu')\phi(R,\mu',t)\mu'd\mu'
\qquad -1 < \mu <0,\end{equation} where $r$ is the distance from
the center of the sphere and $\mu$ is the cosine of the angle the
particle velocity makes with the radius vector, i.e. $(r,\mu) \in
[0,R] \times [-1,1].$ The operator $\mathcal{K}$ is a bounded
positive operator in $X_p=L^p([0,R] \times [-1,1], r^2drd\mu)
\quad (1 < p < \infty).$ We make the general assumptions:
\begin{enumerate}[i)\:]
\item The collision frequency $\Sigma(\cdot,\cdot)$ is bounded and
nonnegative on $[0,R] \times [-1,1].$ \item The kernel
$\kappa(\cdot,\cdot)$ is nonnegative and such that the mapping
$$\mathbf{K}\;:\;f \mapsto
\int_0^1\kappa(\mu,\mu')f(\mu')\mu'd\mu' \:\in
\mathfrak{C}(L^p([-1,0],|\mu|d\mu),L^p([0,1],|\mu|d\mu)).$$ \item
The reflective coefficient $\gamma(\cdot)$ is measurable and $0
\leq \gamma(\mu)\leq \gamma_0 <1.$ \item The collision operator
$\mathcal{K} \in \mathfrak{B}(X_p)$ is regular $(1 < p < \infty)$.
\end{enumerate}
Define the boundary operator $\mathbf{H}=\mathbf{J}+\mathbf{K}$
where
$$\mathbf{J}f(\mu)=\gamma(\mu)f(-\mu) \qquad \forall \mu \in (0,1),\:f \in
L^p([-1,0],|\mu|d\mu)$$ and the transport operator $\textbf{A}_{\H}$
by
$$\textbf{A}_{\H}\phi(r,\mu)=-\mu \frac{\partial
\phi}{\partial r}(r,\mu)-\frac{1-\mu^2}{r}\frac{\partial
\phi}{\partial \mu}(r,\mu)-\Sigma(r,\mu)\phi(r,\mu)$$ with domain
$\D(\textbf{A}_{\H})$ equals to
\begin{equation*}\{\phi \in X_p \text{ such that }
\textbf{A}_{\H}\phi \in X_p, \phi(R,\mu) \in L^p([-1,0],|\mu|d\mu)
\text{ and satisfies } \eqref{sphbc}\}.\end{equation*} The main
properties of the transport operator $\textbf{A}_{\H}$ for various
boundary operator $\H$ has been dealt with in \cite{zhang, zhang1}
and it spectrum has been described in full generality in \cite{ttsp}
for Maxwell--like boundary operator $\H$ satisfying assumptions
ii)--iii). In particular, a consequence of \cite[Theorem 6.8]{gener}
is the following generation result:
\begin{theo}\label{gensph} Assume $i)-iv)$ to be fulfilled.
Then, $\mathbf{A}_{\H}$ generates a nonnegative $c_0$--semigroup
$\ut$ of $X_p$ $(1 < p < \infty)$. As a consequence,
$\textbf{A}_{\H}+\mathcal{K}$ is also the generator of a
$c_0$--semigroup $\vt$ of $X_p$.\end{theo}
\begin{nb} Let us say a few words about the proof of Theorem \ref{gensph}.
As it is well--known \cite{ttsp} (see also Section below), up to a
suitable change of variables, $\mathbf{A_H}$ is similar to a
one--velocity transport operator $T_H$ acting on some Banach space
$\mathcal{X}_p$ (see \eqref{T_H} below for details). Under
assumptions $i)-iv)$, it is then a direct consequence of
\cite[Theorem 6.8]{gener} that $T_H$ generates a $c_0$--semigroup
in $\mathcal{X}_p$ $(1 < p < \infty)$. This implies obviously that
$\mathbf{A}_{\H}$ is a generator of a $c_0$--semigroup in
$X_p$.\end{nb} The main result of this section is then the
following:
\begin{theo}\label{main2} Let $1 < p < \infty$ and let $\mathcal{K} \in \mathfrak{B}(X_p)$
be regular. Then $V(t)-U(t)$ is compact for any $t \geq 0$ and
$\sigma_{\mathrm{ess}}(U(t))=\sigma_{\mathrm{ess}}(V(t))$ $\forall
t \geq 0.$ Moreover,
$\sigma_{\mathrm{ess}}(\textbf{A}_{\H}+\mathcal{K})=\sigma_{\mathrm{ess}}(\textbf{A}_{\H})=\sigma(\textbf{A}_{\H}).$
\end{theo}
\begin{nb} A very precise description of $\sigma(\textbf{A}_{\H})$ can be found in \cite{ttsp}. \end{nb}

\subsection{Proof of Theorem \ref{main2}}

The method of the proof is very similar to that used in the proof
of Theorem \ref{main} and consists in applying Theorem
\ref{sbihi}. We resume briefly some of the arguments developed in
Section \ref{secmain}. Define $R_1(t)=V(t)-U(t)$ for any $t \geq
0$. The proof consists in proving that $R_1(t) \in
\mathfrak{C}(X_p)$ for any $t \geq 0$. Since $\mathcal{K}$ is a
nonnegative operator, it is easy to see that it suffices to prove
the result for a constant collision frequency, say \bq
\label{const} \Sigma(r,\mu)=\Sigma \qquad \text{ for any } (r,\mu)
\in [0,R] \times [-1,1].\eq 
Moreover, one may assume without loss of generality that
$$\|\mathbf{H}\| < 1.$$ As above, since $\mathcal{K}$ is regular
and $\mathbf{K}$ is compact, it suffices to prove the result for a
collision operator of the form \bq \label{coll}
\mathcal{K}\varphi(r,\mu)=\sum_{i \in
I}\alpha_i(r)\beta_i(\mu)\int_{-1}^1\theta_i(\mu_{\star})\varphi(r,\mu_{\star})\d
\mu_{\star}\eq where $I \subset \mathbb{N}$ is finite,
$\alpha_i(\cdot) \in L^{\infty}([0,R])$ and
$\beta_i(\cdot),\,\theta_i(\cdot) \in \mathcal{C}_c(]-1,1[)$ $(i
\in I)$ and for a kernel $\kappa(\cdot,\cdot)$ which reads \bq
\label{kappa} \kappa(\mu,\mu')=\sum_{j \in
J}\mathbf{g}_j(\mu)\mathbf{k}_j(\mu') \eq where $J \subset
\mathbb{N}$ is finite, $\mathbf{g}_j(\cdot) \in
\mathcal{C}_c(]0,1[)$ and $\mathbf{k}_j \in
\mathcal{C}_c(]-1,0[)$, $j \in J.$ In such a case, $R_1(t) \in
\bigcap_{1 < r < \infty} \mathfrak{B}(X_p)$ so that it suffices to
prove the compactness of $R_1(t)$ for $p=2.$ Throughout the
sequel, we will therefore restrict ourselves to the case $p=2$ and
will assume \eqref{const}, \eqref{coll} and \eqref{kappa} to be
satisfied. At this point it is convenient to use a change of
variable already performed in \cite{ttsp} (see also
\cite{sahni1,zhang,zhang1}): let $x=r\mu$ and $y=r\sqrt{1-\mu^2}.$
This transformation is one--to--one from $[0,R] \times [-1,1]$
onto $\Omega=\{(x,y);\,x^2+y^2 \leq R, 0 \leq y \leq R\}.$ Then,
there exists an isometric isomorphism $\mathcal{J}$ from $X_2$ to
$\mathcal{X}_2:=L^2(\Omega,ydydx)$ defined as
\begin{equation*}\begin{cases}
\mathcal{J}\::\:&L^2([0,R] \times [-1,1],r^2drd\mu) \to \mathcal{X}_2\\
&\phi(r,\mu) \mapsto \mathcal{J}\phi(x,y)=\phi(\sqx,x/\sqx), \qquad
(x,y) \in \Omega.\end{cases}\end{equation*} In this case, the
transport operator $\mathbf{A_H}=\mathcal{J}^{-1}T_H\mathcal{J}$
where $T_H$ is the following transport operator:
\begin{equation}\label{T_H}\begin{cases}
T_H\::\: &\D(T_H) \to \mathcal{X}_2\\
&\varphi \mapsto T_H\varphi(x,y)=-\dfrac{\partial \varphi}{\partial
x}(x,y)-\Sigma \varphi(x,y),
\end{cases}\end{equation}
whose domain $\D(T_H)$ is
$$\left\{\psi \in \mathcal{X}_2 \text{ such that }
T_H\varphi \in \mathcal{X}_2; \psi(y_{\pm},y) \in Y_2 \text{ and }
\psi(y_-,y)=H\,\psi(y_+,y) \right\}$$ where $y_{\pm}=\pm \sqr$ $(y
\in S=[0,R])$ and $Y_2=L^2(S,y\d y).$ The boundary operator $H \in
\mathfrak{B}(Y_2)$ is given by $H=J+K$ where
$J\varphi(y)=\alpha(y)\varphi(y)$ and
$$K\varphi(-\sqr, y)=\int_0^Rk(y,y')\varphi(-\sqrt{R^2-y^{'2}},y')\dfrac{y'}{R}dy' \qquad (y \in S)$$ where the new scattering kernel $k(\cdot,\cdot)$ and the
reflective coefficient $\alpha(\cdot)$ are defined as
$$k(y,y')=\kappa(-\sqr /R,\sqr /R ),\qquad \alpha(y)= \gamma(-\sqr/R).$$
Note that, since $\|\mathbf{H}\| < 1$ and $\mathcal{J}$ is
isometric, one has $\|H\| < 1$ so that $T_H$ is dissipative.

In the same way, one can define the following collision operator
$\mathcal{B}=\mathcal{JKJ}^{-1} \in \mathfrak{B}(\mathcal{X}_2)$.
Straightforward computations yield
\begin{multline}\label{matB}
 \mathcal{B}\varphi(x,y)=\sum_{i \in
I}\alpha_i(\sqx)\beta_i(x /\sqx)\times
\\
\times \int_{-\sqx}^{\sqx}\theta_i(z/\sqx
)\varphi(z,\sqrt{x^2+y^2-z^2})\dfrac{\d z}{\sqx}.\end{multline}

Let us denote by $(\mathcal{U}(t))_{t \geq 0}$ and
$(\mathcal{V}(t))_{t \geq 0}$ the $c_0$--semigroups in
$\mathcal{X}_2$ generated by $T_H$ and $T_H+\mathcal{B}$
respectively. Since
$R_1(t)=\mathcal{J}^{-1}(\mathcal{V}(t)-\mathcal{U}(t))\mathcal{J}$
for any $t \geq 0$, one has to prove that
$\mathcal{V}(t)-\mathcal{U}(t) \in \mathfrak{C}(\mathcal{X}_2)$.
To this aim we shall apply Theorem \ref{sbihi} and we have to
compute explicitly the resolvent of $T_H$. Define for any
$\mathrm{Re}\lambda
> -\Sigma$:\begin{equation*}\begin{cases}
M_{\lambda}:&\: Y_2 \longrightarrow Y_2\\
&u \longmapsto
M_{\lambda}u(y)=u(y)\exp\{-2(\lambda+\Sigma)\sqr\},\quad (y \in
S)\end{cases}
\end{equation*}
\begin{equation*}\begin{cases}
\Xi_{\lambda}:&\: Y_2 \longrightarrow \mathcal{X}_2\\
&u \longmapsto
\Xi_{\lambda}u(x,y)=u(y)\exp\{-(\lambda+\Sigma)(x+\sqr)\},\end{cases}
\end{equation*}

\begin{equation*}\begin{cases}
G_{\lambda}:&\mathcal{X}_2 \longrightarrow Y_2\\
&\varphi \longmapsto
G_{\lambda}\varphi(y)=\displaystyle\int_{-\sqr}^{\sqr}\varphi(z,y)e^{-(\lambda+\Sigma)(\sqr-z)}\d
z\end{cases}
\end{equation*}
and\begin{equation*}\begin{cases}
C_{\lambda}:&\: \mathcal{X}_2 \longrightarrow \mathcal{X}_2\\
&\varphi \longmapsto C_{\lambda}\varphi(x,y)=\displaystyle
\int_{-\sqr}^{x}\varphi(z,y)e^{-(\lambda+\Sigma)(x-z)}\d
z.\end{cases}
\end{equation*}
The resolvent of $T_H$ is then given by the following (see
\cite{ttsp})
\begin{propo}\label{reso}
Let $H \in \mathfrak{B}(Y_2)$ be given as above. Then
\begin{equation*}\label{eqres}
(\lambda-T_H)^{-1}=\Xi_{\lambda}H(I-M_{\lambda}H)^{-1}G_{\lambda}+C_{\lambda}\qquad
\forall \,Re\lambda > -\Sigma.\end{equation*}
\end{propo}
Then, the key point of the proof of Theorem \ref{main2} stands in
the following whose proof is given in the Appendix \ref{app2}:
\begin{propo}\label{iml2} For any regular operator
$B \in \mathfrak{B}(\mathcal{X}_2)$ and any $\mathrm{Re}\lambda >
-\Sigma$:
$$\lim_{|\mathrm{Im}\lambda| \to \infty}
\left(\|B^{\star}(\lambda-T_H)^{-1}B\|_{\mathfrak{B}(\mathcal{X}_2)}
+\|B(\lambda-T_H)^{-1}B^{\star}\|_{\mathfrak{B}(\mathcal{X}_2)}\right)=0
.$$
\end{propo}

\begin{proof2}  The proof of Theorem \ref{main2} is now a
straightforward application of Theorem \ref{sbihi} as in Theorem
\ref{main}.\end{proof2}

\section{Transport equations in slab geometry}\label{VIpreli}
%
%
%
Let us consider the following transport equation in a slab with
thickness $2a>0$:
\begin{equation}\label{slab1}
\dfrac{\partial \varphi}{\partial t}(x,\xi,t)+\xi \dfrac{\partial
\varphi}{\partial
x}(x,\xi,t)+\sigma(x,\xi)\varphi(x,\xi,t)=\int_{-1}^1\kappa(x,\xi,\xi_{\star})\varphi(x,\xs,t)\d\xs
\end{equation}
with the boundary conditions
\begin{equation}\label{slabbc}
\varphi^i=H(\varphi^o)
\end{equation}
and the initial datum $\varphi(x,\xi,t=0)=\phi_0(x,\xi) \in
X_p=L^p([-a,a] \times [-1,1];\d x\d \xi)$ $(1 < p< \infty).$ The
incoming boundary of the phase space $D^i$ and the outgoing one
$D^o$ are given by :
$$D^i:=D^i_1 \cup D^i_2:=\{-a\} \times [0,1] \cup \{a\} \times [-1,0],$$
$$D^o:=D^o_1 \cup D^o_2:=\{-a\} \times [-1,0] \cup \{a\} \times [0,1],$$
while the associated boundary spaces are
$$X^i_p:=L^p(D^i_1,|\xi|d\xi) \times L^p(D^i_2,|\xi|d\xi)=
X^i_{1,\,p} \times X^i_{2,\,p},$$ and $$
X^o_p:=L^p(D^o_1,|\xi|d\xi) \times L^p(D^o_2,|\xi|d\xi)=
X^o_{1,\,p} \times X^o_{2,\,p},$$ endowed with their natural norms
(see \cite{lalods} for details).
Let $W_p$ be the partial Sobolev space $W_p:=\{\psi \in X_p
\:\mbox{ such that } \xi  \frac{\partial \psi} {\partial x}\in
X_p\}.$ Any function $\psi \in W_p$ admits traces on $D^o$ and
$D^i$ denoted by $\psi^o$ and $\psi^i$ respectively. Precisely,
$\psi^o=(\psi^o_1,\psi^o_2)$ and $\psi^i=(\psi^i_1,\psi^i_2)$ are
given by
\begin{eqnarray}
\label{VIpsiopsii}
\begin{cases}
\begin{array}{lcl}
\psi^o_1(\xi)=\psi(-a,\xi) &&\xi \in (-1,0);\\
\psi^o_2(\xi)=\psi(a,\xi)  &&\xi \in (0,1);\\
\psi^i_1(\xi)=\psi(-a,\xi) &&\xi \in (0,1);\\
\psi^i_2(\xi)=\psi(a,\xi) &&\xi \in (-1,0).\\
\end{array}
\end{cases}\end{eqnarray}
We describe the boundary operator $H$ relating the incoming flux
$\psi^i$ to the outgoing one $\psi^o$ by
\begin{equation*}
\begin{cases}
H\::X^o_{1,\,p} \times X^o_{2,\,p} \: \to X^i_{1,\,p} \times X^i_{2,\,p}\\
H \left(\begin{array}{c} u_1 \\ u_2 \end{array} \right):=
\left(\begin{array}{cc} H_{11} & H_{12}\\
H_{21} & H_{22}
\end{array}
\right) \left(\begin{array}{c} u_1 \\ u_2 \end{array} \right)
\end{cases}
\end{equation*}
where  $H_{jk} \in \mathcal{L}(X^o_{k,p};X^i_{j,p});j,k=1,2.$ Let us
now define the transport operator associated to the boundary
conditions induced by $H$
\begin{equation*}
\begin{cases}
\begin{split}
T_H\; :
\; &\D(T_H) \subset X_p \to X_p\\
&\psi \mapsto T_H\psi(x,\xi)=-\xi\frac{\partial \psi}{\partial x}(x,\xi)-\sigma(x,\xi)\psi(x,\xi),\\
\end{split}
\end{cases}
\end{equation*}
where
$$\D(T_H)=\{\psi \in W_p \mbox{ such that } \psi^o \in X^o_p \text{ and } H\psi^o=\psi^i\}.$$

We make the following assumptions

$(\mathrm{H}1)$ The collision frequency $\sigma(\cdot,\cdot)$ is
measurable, bounded and nonnegative on $[-a,a] \times [-1,1].$

$(\mathrm{H}2)$ The collision kernel $\kappa(\cdot,\cdot,\cdot)$
is measurable and nonnegative on $[-a,a] \times [-1,1] \times
[-1,1]$ and such that the operator
$$\mathcal{K}\;:\;f(x,\xi) \mapsto
\mathcal{K}f(x,\xi)=\int_{-1}^1\kappa(x,\xi,\xs)f(x,\xs)\d\xs $$
is \textit{regular} on $ X_p$ $(1 < p < \infty)$.\medskip

Concerning the boundary operator $H$ we assume that one of the
following assumptions is fulfilled:
\begin{enumerate}[a)\:]
\item $H$ is a diagonal operator of the form $H=\left(\begin{array}{cc} H_{11} & 0\\
0 & H_{22}
\end{array}
\right)$ with $H_{11}=\rho_1J_1+ K_1$ and $H_{22}=\rho_{2}J_2+K_2$
where $\rho_i$ is positive $(i=1,2)$ and $K_i$ is a compact
operator. The operators $J_i$ $(i=1,2)$ are given by
\begin{equation*}\begin{cases}
J_1\::&X^0_{1,p} \to X^i_{1,p} \\
&\psi(-a,\cdot) \mapsto
J_1\psi(\xi)=\psi(-a,-\xi)\end{cases}\end{equation*}
\begin{equation*}\begin{cases}
J_2\::&X^0_{2,p} \to X^i_{2,p} \\
&\psi(a,\cdot) \mapsto
J_2\psi(\xi)=\psi(a,-\xi)\end{cases}\end{equation*}
\item $H$ is a off--diagonal operator of the form $H=\left(\begin{array}{cc} 0 & H_{12}\\
H_{21} & 0
\end{array}
\right)$ with $H_{12}=\beta_1I_{12}+ K_1$ and
$H_{21}=\beta_{2}I_{21}+K_2$ where $\beta_i$ is positive $i=1,2$ and
$K_i$ is a compact operator. The operators $I_{12}$ and $I_{21}$ are
given by
\begin{equation*}\begin{cases}
I_{12}\::&X^0_{2,p} \to X^i_{1,p} \\
&\psi(a,\cdot) \mapsto
I_{12}\psi(\xi)=\psi(-a,\xi)\end{cases}\end{equation*}
\begin{equation*}\begin{cases}
I_{21}\::&X^0_{1,p} \to X^i_{2,p} \\
&\psi(-a,\cdot) \mapsto
I_{21}\psi(\xi)=\psi(a,\xi)\end{cases}\end{equation*} \item The
boundary operator $H$ is compact.\end{enumerate}

There is a vast literature dealing with model
\eqref{slab1}--\eqref{slabbc} starting with the pioneering work of
Lehner and Wing \cite{lehner}. We only mention the recent results
of \cite{lalods} dealing with the asymptotic behavior of the
solution to \eqref{slab1}--\eqref{slabbc} as well as \cite{
chabi1, chabi2} which take into account possibly unbounded
collision operator $\mathcal{K}$. In the $L^p$--setting, our main
result generalizes the existing ones:
\begin{theo}\label{slab} Let $1 < p < \infty.$
Let $(\mathrm{H1})$ and $(\mathrm{H2})$ be fulfilled. Moreover,
assume that $H$ satisfies one of the assumptions a), b) or c).
Then, $V(t)-U(t) \in \mathfrak{C}(X_p)$ for any $t \geq 0$. In
particular,
$\sigma_{\mathrm{ess}}(V(t))=\sigma_{\mathrm{ess}}(U(t))$ $(t \geq
0)$ where $\vt$ is the $c_0$--semigroup generated by
$T_H+\mathcal{K}$ and $\ut$ is the one generated by
$T_H$.\end{theo}

\begin{preuve} Note that the existence of the semigroup $\ut$
generated by $T_H$ is a direct consequence of \cite{gener}. To
prove that $V(t)-U(t) \in \mathfrak{C}(X_p)$ one sees easily,
arguing as above that it suffices to prove the result for $p=2$,
$\|H\| < 1$ and a constant collision frequency
$\sigma(x,\xi)=\sigma.$ In this case, one deduces from
\cite[Theorem 2.1, p. 55]{lathese}, \cite[Proposition
3.1]{lalods}, and \cite[Theorem 3.2, p. 77]{lathese} that property
\eqref{imlam} of Theorem \ref{sbihi} holds. The previous
references correspond respectively to the assumption a), b) and
c). Since $\mathcal{K}(\lambda-T_H)^{-1}$ is compact for
$\mathrm{Re}\lambda > -\sigma$ \cite{latrach}, one concludes
thanks to Theorem \ref{sbihi}.\end{preuve}
\begin{nb} Note that, in \cite{dehici} (see also
\cite{chabi1,chabi2}) the identity
$r_{\mathrm{ess}}(V(t))=r_{\mathrm{ess}}(U(t))$ is established
exploiting the explicit nature of $\ut$ in the case of perfect
reflecting boundary conditions or periodic conditions. Even if,
for general boundary operator $H$ satisfying a)--c) the semigroup
$\ut$ can also be made explicit (see for instance \cite{these}),
the resolvent approach is much more easy to apply and leads to
similar results.
\end{nb}


\section{Appendix 1: Proof of Proposition \ref{iml}}\label{app1}

The aim of this Appendix is to prove the Proposition \ref{iml}. We
decompose its proof into several steps. The strategy is inspired
by similar results in \cite{lalods}. First, since $B$ is of the
form \eqref{regular}, it is enough to show by linearity that
$$\lim_{|\mathrm{Im}\lambda| \to \infty}
\|B_1(\lambda-\mathbf{A}_H)^{-1}B_2\|=0 \qquad \forall \:
\mathrm{Re}\lambda
> -\underline{\sigma}$$
where
$$B_i\varphi(\mu,v)=\alpha_i(\mu)\beta_i(v)\int_a^b
\theta_i(v_{\star})\varphi(\mu,v_{\star})\d v_{\star},\qquad
(i=1,2)$$ and $\alpha_i(\cdot) \in L^{\infty}(]0,1[$,
$\beta_i(\cdot), \theta_i(\cdot) \in \mathcal{C}_c(]a,b[)$
$i=1,2.$ This shall be done in several steps. Recall that, by
Proposition \ref{reso},
$$(\lambda-\mathbf{A}_H)^{-1}=\Xi_{\lambda}H(I-M_{\lambda}H)^{-1}G_{\lambda}+C_{\lambda}\qquad
Re\lambda > -\underline{\sigma}.$$

\noindent {\bf Step 1:} We first note that, for any
$\mathrm{Re}\lambda> -\underline{\sigma}$ the operator
$C_{\lambda}$ is nothing else but the resolvent of the transport
operator $\mathbf{A}_H$ in the case of absorbing boundary
conditions, $H=0$. Then, according to a result by M.
Mokhtar--Kharroubi \cite[Lemma 2.1]{VImmklaplace},
$$\lim_{|\mathrm{Im}\lambda| \to \infty}\|B_1C_{\lambda}B_1\|=0 \qquad
\forall\:\mathrm{Re}\lambda > - \underline{\sigma}.$$
 Therefore, it is enough to prove that
\bq \label{aH-A0} \lim_{|\mathrm{Im}\lambda| \to
\infty}\|B_1\Xi_{\lambda}H(I-M_{\lambda}H)^{-1}G_{\lambda}B_2\|=0
\qquad \forall\:\mathrm{Re}\lambda > - \underline{\sigma}.\eq

\noindent {\bf Step 2:} We note that, adapting the result of
\cite[Theorem 3.2, p. 77]{lathese} (see \cite{jeribi} for
details), one has \bq \label{aK} \lim_{|\mathrm{Im}\lambda| \to
\infty}\|B_1\Xi_{\lambda}K(I-M_{\lambda}H)^{-1}G_{\lambda}B_2\|=0
\qquad \forall\:\mathrm{Re}\lambda = - \underline{\sigma}+\omega,
\omega > 0.\eq

\noindent {\bf Step 3.} Using the fact that $H=\beta \mathrm{Id}
+K$, it remains only to show that $$ \lim_{|\mathrm{Im}\lambda| \to
\infty}\|B_1\Xi_{\lambda}(I-M_{\lambda}H)^{-1}G_{\lambda}B_2\|=0
\qquad \forall\:\mathrm{Re}\lambda = - \underline{\sigma}+\omega,
\omega > 0.$$ To do, using the fact that
$(I-M_{\lambda}H)^{-1}=\sum_{n=0}^{\infty} (M_{\lambda}H)^n,$
together with the dominated convergence theorem, it suffices to show
that, for any integer $n \in \mathbb{N}$ \bq \label{an}
\lim_{|\mathrm{Im}\lambda| \to
\infty}\|B_1\Xi_{\lambda}(M_{\lambda}H)^{n}G_{\lambda}B_2\|=0 \qquad
\forall\:\mathrm{Re}\lambda = - \underline{\sigma}+\omega, \omega >
0.\eq Since $M_{\lambda}H=\beta M_{\lambda}+M_{\lambda}K$, for any
$n \in \mathbb{N}$, $(M_{\lambda}H)^n=\sum_{j=1}^{2^n}P_j(\lambda)$
where $P_j(\lambda)$ is the product of $n$ factors formed with
$\beta M_{\lambda}$ and $M_{\lambda}K$. Among these factors, only
$P_{2^n}(\lambda)=(\beta M_{\lambda})^n$ does not involve $K$
whereas, for $j \in \{1,\ldots,2^n-1\}$, the operator $K$ appears at
least once in the expression of $P_j(\lambda)$.\\

\noindent {\bf Step 3.1 :} One proves that, for any $j \in
\{1,\ldots,2^n-1\}$,
$$\lim_{|\mathrm{Im}\lambda| \to
\infty}\|P_j(\lambda)G_{\lambda}B_2\|_{\mathfrak{B}(X_2,Y_2)}=0
\qquad \forall \mathrm{Re}\lambda= -\underline{\sigma} + \omega,
\omega > 0.$$ By assumption, there exists $k \in \{0,\ldots, n-1\}$
such that $P_j(\lambda)=P^1_j(\lambda)M_{\lambda}K(\beta
M_{\lambda})^k$ where $P^1_j(\lambda)$ is a product of operators
$M_{\lambda}K$ and $\beta M_{\lambda}.$ As a by--product,
$$\sup \{\|P^1_j(\lambda)M_{\lambda}\|\,;\,\mathrm{Re}\lambda =
-\underline{\sigma}+ \omega\} < \infty.$$ It suffices then to show
that, for any $k \geq 0$, \bq \label{mjk} \lim_{|\mathrm{Im}\lambda|
\to \infty}\|K(\beta M_{\lambda})^k
G_{\lambda}B_2\|_{\mathfrak{B}(X_2,Y_2)}=0 \qquad \forall
\mathrm{Re}\lambda > -\underline{\sigma} + \omega.\eq A direct
computation shows that
\begin{equation*}\begin{split}
KM_{\lambda}^k G_{\lambda}B_2\varphi(v)=\dfrac{\alpha}{v}\int_a^b
k(v,\v)\beta_2(\v)\exp\{-k(\lambda+
\underline{\sigma})/\v\}\d \v \times\\
\times\int_0^1\exp\{-
\dfrac{(1-\mu')}{\v}(\lambda+\underline{\sigma})\}\alpha_2(\mu')\d
\mu' \int_a^b \theta_2(w)\varphi(\mu',w)\d
w.\end{split}\end{equation*} Then, one may decompose $KM_{\lambda}^k
G_{\lambda}B_2$ as $KM_{\lambda}^k
G_{\lambda}B_2=\mathcal{R}_1(\lambda)\mathcal{R}_2$ with
$$\mathcal{R}_2 \::\varphi \in X_2 \mapsto
\mathcal{R}_2\varphi(\mu)=\alpha_2(\mu)\int_a^b
\theta_2(w)\varphi(\mu,w)\d w \in L^2(]0,1[,\d \mu)$$ and
\begin{equation*}\begin{split}
\mathcal{R}_1(\lambda)\psi(v)=\dfrac{\alpha}{v}\int_a^b
k(v,\v)\beta_2(\v)\exp\{-k(\lambda+
\underline{\sigma})/\v\}\d \v \times\\
\times\int_0^1\exp\{-
\dfrac{(1-\mu')}{\v}(\lambda+\underline{\sigma})\}\psi(\mu')\d \mu'
\in X_2, \qquad  \psi \in L^2(]0,1[,\d \mu).
\end{split}\end{equation*}
It is then enough to show that $\lim_{|\mathrm{Im}\lambda| \to
\infty}\|\mathcal{R}_1(\lambda)\|=0$ for any
Re$\lambda=-\underline{\sigma}+\omega.$ By linearity, using that
the kernel $k(v,\v)$ is of the form \eqref{kdegen}, one may assume
without loss of generality that $k(v,\v)=g(v)k(\v)$ where both
$g(\cdot)$ and $k(\cdot)$ are continuous functions with compact
supports in $]a,b[$. Now, let us fix $\psi \in L^2(]0,1[,\d \mu)$
and denote by $\widetilde{\psi}$ its trivial extension to
$\mathbb{R}$. Then, one sees easily that
$$\mathcal{R}_1(\lambda)\psi(v)=\dfrac{g(v)}{v}\int_{\mathbb{R}}F_{\lambda}(k+1-\mu')\widetilde{\psi}(\mu')\d
\mu'$$ where
$$F_{\lambda}(x)=\alpha \int_a^b
k(\v)\beta_2(\v)\exp\{-
\dfrac{x}{\v}(\lambda+\underline{\sigma})\}\d \v \qquad \forall x
\geq 0.$$ 
One sees that \begin{equation*}
\label{uniF}\int_0^{\infty}\sup_{\mathrm{Re}\lambda=-\underline{\sigma}+\omega}|F_{\lambda}(x)|^2
\d x \leq \dfrac{\alpha^2}{2\omega}\int_a^b |k(\v)|^2 \v\, \d \v
\int_a^b |\beta_2(\v)|^2\d \v < \infty.\end{equation*} According
to Riemann--Lebesgue Lemma \ref{riemann} and the Dominated
Convergence Theorem, it is not difficult to see that
$$\lim_{|\mathrm{Im}\lambda|\to \infty}
\int_0^{\infty}|F_{\lambda}(x)|^2 \d x=0 \qquad \forall
\mathrm{Re}\lambda=-\underline{\sigma}+\omega.$$ Since
$$\|\mathcal{R}_1(\lambda)\|^2 \leq \left(\int_a^b \left|\dfrac{g(v)}{v}\right|^2 v \d v\right) \int_0^{\infty}|F_{\lambda}(x)|^2 \d x.$$
this proves the desired result. It remains to investigate the case $j=2^n$:\\

\noindent {\bf Step 3.2 :} It remains to evaluate the behavior of
$\|B_1\Xi_{\lambda}P_{2^n}G_{\lambda}B_2\|$ as
$|\mathrm{Im}\lambda|$ goes to infinity, where $P_{2^n}=(\beta
M_{\lambda})^{n}$. Precisely, let us show that \bq \label{anbeta}
\lim_{|\mathrm{Im}\lambda| \to \infty}\|B_1\Xi_{\lambda}(\beta
M_{\lambda})^{n}G_{\lambda}B_2\|=0 \qquad
\forall\:\mathrm{Re}\lambda = - \underline{\sigma}+\omega, \omega
> 0.\eq
Let $\varphi \in X_2$. Straightforward calculations  yield
\begin{equation*}\begin{split}
B_1\Xi_{\lambda}M_{\lambda}^{n}&G_{\lambda}B_2\varphi(\mu,v)=\alpha_1(\mu)\beta_1(v)
\int_a^b\frac{\theta_1(\v)\beta_2(\v)}{\v}\exp\{-\dfrac{(\lambda+\underline{\sigma}\,)}{\v}\mu\}\d
\v \times\\
&\times
\int_0^1\alpha_2(\mu')\exp\{-\dfrac{(\lambda+\underline{\sigma}\,)}{\v}(n+1-\mu')\}\d
\mu' \int_a^b\theta_2(w)\varphi(\mu',w)\d w.\end{split}
\end{equation*}
As above, one may split this operator as
$B_1\Xi_{\lambda}M_{\lambda}^{n}G_{\lambda}B_2=\mathcal{A}_3\mathcal{A}_2(\lambda)\mathcal{A}_1$
where
$$\mathcal{A}_1:\varphi \in X_2 \mapsto
\mathcal{A}_1\varphi(\mu)=\alpha_2(\mu)\int_a^b\theta_2(w)\varphi(\mu,w)\d
w \in L^2(]0,1[,\d \mu),$$
\begin{equation*}\begin{cases}
\mathcal{A}_2(\lambda)\::\:L^2(]0,1[,\d \mu) \to L^2(]0,1[,\d
\mu)\\
\psi \mapsto \mathcal{A}_2(\lambda)\psi(\mu)=\displaystyle\int_a^b\d
\v
\int_0^1\frac{\theta_1(\v)\beta_2(\v)}{\v}\psi(\mu')e^{-\frac{(\lambda+\underline{\sigma}\,)}{\v}
(n+1+\mu-\mu')}\d \mu'
\end{cases}
\end{equation*} and
$$\mathcal{A}_3\::\:\psi \in L^2(]0,1[,\d \mu) \mapsto
\alpha_1(\mu)\beta_1(v)\psi(\mu) \in X_2.$$ It is clearly sufficient
to prove that
$$\lim_{|\mathrm{Im}\lambda| \to
\infty}\|\mathcal{A}_2(\lambda)\|=0 \qquad \forall
\;\mathrm{Re}\lambda=-\underline{\sigma}+\omega.$$ As in the proof
of Step 3.1, let us define, for any $x \in \mathbb{R}$
$$F_{\lambda}(x)=\int_a^b \frac{\theta_1(\v)\beta_2(\v)}{\v}
\exp\{-\frac{(\lambda+\underline{\sigma}\,)}{\v} x\}\d  \v$$ so that
$$\mathcal{A}_2(\lambda)\psi(\mu)=\int_{\mathbb{R}}F_{\lambda}(n+1+\mu-x)\widetilde{\psi}(x) \d
x, \qquad \psi \in L^2(]0,1[,\d \mu)$$ where $\widetilde{\psi}$ is
the trivial extension to $\mathbb{R}$ of $\psi \in L^2(]0,1[,\d
\mu).$ As in the proof of Step 3.1, one can show that
$$\int_{\mathbb{R}}\left(\sup_{\mathrm{Re}\lambda=-\underline{\sigma} +
\omega}\left|F_{\lambda}(x)\right|^2\right) \d x < \infty$$ and
$$\|\mathcal{A}_2(\lambda)\| \leq
\|F_{\lambda}(\cdot)\|_{L^2(\mathbb{R})} \qquad
(\mathrm{Re}\lambda=-\underline{\sigma} + \omega).$$ Then,
applying again Riemmann--Lebesgue Lemma \ref{riemann} together
with the dominated convergence theorem, one gets
$$\lim_{|\mathrm{Im}\lambda|\to \infty}
\|F_{\lambda}(\cdot)\|_{L^2(\mathbb{R})}=0 \qquad
(\mathrm{Re}\lambda=-\underline{\sigma} + \omega),$$ which leads to
the conclusion. Combining all the above steps, we proved Proposition
\ref{iml}.

\section{Appendix 2: Proof of Proposition \ref{iml2}}\label{app2}

In this appendix, we prove Proposition \ref{iml2} which is the key
point of the proof of Theorem \ref{main2}. Since $\mathcal{B}$ is
of the form \eqref{matB}, by a linearity argument it suffices to
prove that, for any $\omega > 0$, \bq\label{B1}
\lim_{|\mathrm{Im}\lambda| \to
\infty}\|\mathcal{B}_1(\lambda-T_H)^{-1}\mathcal{B}_2\|_{\mathfrak{B}(\mathcal{X}_2)}=0
\qquad \forall\,\mathrm{Re}\lambda=-\Sigma + \omega\eq where
\begin{multline*}
\mathcal{B}_i\varphi(x,y)=\dfrac{\alpha_i(\sqx)}{\sqx}
\beta_i(x /\sqx)\times
\\
\times
\int_{-\sqx}^{\sqx}\theta_i(z/\sqx)\varphi(z,\sqrt{x^2+y^2-z^2})\d
z\end{multline*} where $\alpha_i(\cdot) \in L^{\infty}([0,R])$,
$\beta_i(\cdot) \in \mathcal{C}_c(]-1,0[)$ and $\theta_i(\cdot)
\in \mathcal{C}_c(]0,1[)$ $(i=1,2)$. We shall prove \eqref{B1} in
several steps.

\noindent {\bf Step 1:} As in the first step of the proof of
Proposition \ref{iml}, it is a direct consequence of
\cite{VImmklaplace} that
$$\lim_{|\mathrm{Im}\lambda| \to \infty}\|\mathcal{B}_1 C_{\lambda}
\mathcal{B}_2\|=0 \qquad \forall \mathrm{Re}\lambda=-\Sigma +
\omega.$$

\noindent {\bf Step 2:} We show now that
$$\lim_{|\mathrm{Im}\lambda| \to
\infty}\|\mathcal{B}_1\Xi_{\lambda}K(I-M_{\lambda}H)^{-1}G_{\lambda}\mathcal{B}_2\|=0
\qquad \forall\:\mathrm{Re}\lambda = - \Sigma+\omega, \omega >
0.$$ Let us first prove that, for any $\varphi \in Y_2$, \bq
\label{strong} \lim_{|\mathrm{Im}\lambda| \to
\infty}\|\mathcal{B}_1\Xi_{\lambda}\varphi\|_{\mathcal{X}_2}=0
\qquad \mathrm{Re}\lambda>-\Sigma.\eq One has, for a. e. $(x,y)
\in \Omega$:\begin{multline*}
\mathcal{B}_1\Xi_{\lambda}\varphi(x,y)=\alpha_1(\sqx)\beta_1(x/\sqx)\int_{-\sqx}^{\sqx}
\theta_1(z/\sqx)\\
\varphi(\sqrt{x^2+y^2-z^2})\exp\{-(\lambda+\Sigma)(z+\sqrt{R^2+z^2-x^2-y^2})\}\dfrac{\d
z}{\sqx},
\end{multline*}
and the Riemann--Lebesgue Lemma implies that, for any
$\mathrm{Re}\lambda =-\Sigma+\omega$,
$$\lim_{|\mathrm{Im}\lambda| \to
\infty}\left|\mathcal{B}_1\Xi_{\lambda}\varphi(x,y)\right|^2=0
\qquad \text{ a.e. } (x,y) \in \Omega.$$ Then, the dominated
convergence theorem leads to \eqref{strong}. Now, let $B$ be the
unit ball of $\mathcal{X}_2$. It is clear that,
$$M:=\sup\{\|(I-M_{\lambda}H)^{-1}G_{\lambda}\mathcal{B}_2\psi\|\,;\,\psi
\in B,\,\mathrm{Re}\lambda =-\Sigma+ \omega\| < \infty$$ i.e,
$$(I-M_{\lambda}H)^{-1}G_{\lambda}\mathcal{B}_2(B) \subset
\{\varphi \in \mathcal{Y}_2\,;\|\varphi\| \leq M\}.$$ Note that
this last set is a bounded subset of $\mathcal{Y}_2$ which is
independent of $\lambda$. The compactness of $K$ together with
\eqref{strong} ensure then that $$\lim_{|\mathrm{Im}\lambda| \to
\infty}\sup_{\varphi \in \mathcal{Y}_2;\|\varphi\| \leq
M}\|\mathcal{B}_1\Xi_{\lambda}K\varphi\|=0$$ which is the desired
result.

\noindent {\bf Step 3:} Let us show now that, for any $n \in
\mathbb{N}$ \bq \label{ans} \lim_{|\mathrm{Im}\lambda| \to
\infty}\|\mathcal{B}_1\Xi_{\lambda}(M_{\lambda}H)^{n}G_{\lambda}\mathcal{B}_2\|=0
\qquad \forall\:\mathrm{Re}\lambda = - \underline{\sigma}+\omega,
\omega > 0.\eq One writes
$(M_{\lambda}H)^n=\sum_{j=1}^{2^n}P_j(\lambda)$ where
$P_j(\lambda)$ is the product of $n$ factors formed with
$M_{\lambda}J$ and $M_{\lambda}K$ and where $P_{2^n}(\lambda)=(
M_{\lambda}J)^n$. For $j \in \{1,\ldots,2^n-1\}$, the operator $K$
appears at least once in the expression of $P_j(\lambda)$.

\noindent {\bf Step 3.1 :} Let us prove that, for any $j \in
\{1,\ldots,2^n-1\}$,
$$\lim_{|\mathrm{Im}\lambda| \to
\infty}\|P_j(\lambda)G_{\lambda}\mathcal{B}_2\|_{\mathfrak{B}(\mathcal{X}_2,Y_2)}=0
\qquad \forall \mathrm{Re}\lambda= -\Sigma + \omega, \omega > 0.$$

The proof is once again inspired to that of Step 3.1 of Appendix
\ref{app1}. As above, it suffices to prove that, for any $k \geq
0$, \bq \label{mjks} \lim_{|\mathrm{Im}\lambda| \to \infty}\|K(
M_{\lambda}J)^k
G_{\lambda}\mathcal{B}_2\|_{\mathfrak{B}(\mathcal{X}_2,Y_2)}=0
\qquad \forall \mathrm{Re}\lambda > -\Sigma + \omega.\eq One may
assume by a domination argument that the reflection coefficient
$\gamma(\cdot)$ is constant and equals to one. Then, direct
computations show that, for any $y \in [0,R]$, $K( M_{\lambda}J)^k
G_{\lambda}\mathcal{B}_2\varphi(y)$ is equal to
\begin{multline*}
\mathbf{g}(-\sqr /R)\int_0^R \mathbf{k}(\sqrt{R^2-\eta^2}/R)
\exp\{-2k(\lambda+\Sigma)\sqry\}\frac{\y}{R}\d \y\\
\times \int_{-\sqry}^{\sqry} \alpha_2(\sqz)\beta_2(z/\sqz)
\exp\{-(\lambda+\Sigma)(\sqry -z)  \}\d z\\
\times
\int_{-\sqz}^{\sqz}\theta_2(u/\sqz)\varphi(u,\sqrt{z^2+\y^2-u^2})\dfrac{\d
u}{\sqz}.\end{multline*} Therefore, $K( M_{\lambda}J)^k
G_{\lambda}\mathcal{B}_2$ splits as $K( M_{\lambda}J)^k
G_{\lambda}\mathcal{B}_2=\mathcal{R}_1(\lambda)\mathcal{R}_2$ where
$\mathcal{R}_2 \in \mathfrak{B}(Y_2,Y_2)$ is given by
\begin{equation*}
\mathcal{R}_2\varphi(\varrho)=\alpha_2(\varrho)\int_{-\varrho}^{\varrho}
\theta_2(u/\varrho)\varphi(u,\sqrt{\varrho^2-u^2})\dfrac{\d
u}{\varrho} \qquad (\varrho \in [0,R])\end{equation*} and
$\mathcal{R}_1(\lambda) \in \mathfrak{B}(Y_2,Y_2)$ given by
\begin{multline*}
\mathcal{R}_1(\lambda)\psi(y)=\mathbf{g}(-\sqr /R) \int_0^R
\mathbf{k}(\sqrt{R^2-\eta^2}/R)\d \y\\
\times \int_{-\sqry}^{\sqry} \beta_2(z/\sqz)\psi(\sqz)\\
\exp\left\{-(\lambda+\Sigma)\left[(2k+1)\sqry -z\right]\right\}\d
z.
\end{multline*}
It is then enough to show that $\lim_{|\mathrm{Im}\lambda| \to
\infty}\|\mathcal{R}_1(\lambda)\|=0$ $\forall
\mathrm{Re}\lambda=-\Sigma+\omega.$ Splitting the last integral on
integrals over $[-\sqry, 0[$ and $[0, \sqry[$ allows to write
$\mathcal{R}_1(\lambda)=\mathcal{R}_1^-(\lambda)+\mathcal{R}_1^+(\lambda)$.
It is then possible to perform the change of variables $z
\mapsto\varrho=\sqz$ which shows that
\begin{equation*}
\mathcal{R}_1^{\pm}(\lambda)\psi(y)=\mathbf{g}(-\sqr /R)
\int_0^R\psi(\varrho)F_{\lambda}^{\pm}(\varrho)\varrho \d \varrho
\end{equation*}
where
\begin{multline*}F_{\lambda}^{\pm}(\varrho)=\pm \int_0^\varrho \mathbf{k}(\sqrt{R^2-\eta^2}/R)
\beta_2(\sqrt{\varrho^2-\eta^2}/\varrho)\times \\
\times \exp\left\{-(\lambda+\Sigma)\left[(2k+1)\sqry \mp
\sqrt{\varrho^2-\eta^2}\right]\right\} \dfrac{\d
\eta}{\sqrt{\varrho^2-\eta^2}}\end{multline*} Now, noting that
$\omega(\eta)=(2k+1)\sqry \mp\sqrt{\varrho^2-\eta^2}$ fulfills the
assumption of Lemma \ref{riemann} for any $\varrho$, one has
$$\lim_{|\mathrm{Im}\lambda| \to \infty} F_{\lambda}^{\pm}(\varrho)=0
\qquad \text{ for a. e. } \varrho \in (0,R).$$ Moreover, one sees
easily that $\int_0^R\sup_{\mathrm{Re}\lambda=-\Sigma+\omega}
|F_{\lambda}^{\pm}(\varrho)|^2 \varrho \,\d \varrho < \infty$ and
the Dominated Convergence Theorem shows that
$$\lim_{|\mathrm{Im}\lambda| \to \infty}
\int_{0}^R \left|F_{\lambda}^{\pm}(\varrho)\right|^2 \varrho\, \d
\varrho=0 \qquad \forall \mathrm{Re}\lambda=-\Sigma+\omega.$$ Since
$\|\mathcal{R}_1^{\pm}(\lambda)\| \leq \|\mathbf{g}(\cdot)\|_{L^2}
\|F_{\lambda}^{\pm}\|_{Y_2}$ one gets the conclusion.\\

\noindent {\bf Step 3.2 :} Let us show that \bq \label{anbeta1}
\lim_{|\mathrm{Im}\lambda| \to \infty}\|\mathcal{B}_1\Xi_{\lambda}(
M_{\lambda}J)^{n}G_{\lambda}\mathcal{B}_2\|=0 \qquad
\forall\:\mathrm{Re}\lambda = - \Sigma+\omega, \omega
> 0.\eq
Tedious calculations show that $\mathcal{B}_1\Xi_{\lambda}(
M_{\lambda}J)^{n}G_{\lambda}\mathcal{B}_2$ splits as
$\mathcal{B}_1\Xi_{\lambda}(
M_{\lambda}J)^{n}G_{\lambda}\mathcal{B}_2=\mathcal{A}_3\mathcal{A}_2(\lambda)\mathcal{R}_2$
where $\mathcal{R}_2$ has been defined in Step 3.1,
$\mathcal{A}_2(\lambda) \in \mathfrak{B}(Y_2)$ given by:
\begin{equation*}\begin{split}
\mathcal{A}_2(\lambda)\psi(\eta)&=\mathcal{A}_2^1(\lambda)\psi(\eta)+\mathcal{A}_2^2(\lambda)\psi(\eta)\\
&=\int_{-\eta}^{\eta}\theta_1(z /\eta) \dfrac{\d z}{\eta}
\int_{-\sqrt{R^2+z^2-\eta^2}}^{0}
\beta_2(\frac{u}{\sqrt{u^2+\eta^2-z^2}})\\
&\times \psi(\sqrt{u^2+\eta^2-z^2})e^{-(\lambda+\Sigma)
\left[(2n+2)\sqrt{R^2+z^2-\eta^2}-u+z\right]}\d u,\\
&+\int_{-\eta}^{\eta}\theta_1(z /\eta) \dfrac{\d z}{\eta}
\int_{0}^{\sqrt{R^2+z^2-\eta^2}}
\beta_2(\frac{u}{\sqrt{u^2+\eta^2-z^2}})\\
&\times \psi(\sqrt{u^2+\eta^2-z^2})e^{-(\lambda+\Sigma)
\left[(2n+2)\sqrt{R^2+z^2-\eta^2}-u+z\right]}\d u,
\end{split}\end{equation*}
and
$\mathcal{A}_3 \varphi(x,y)=
\alpha_1(\sqx)\beta_1(x/\sqx)\varphi(\sqx) \in \mathcal{X}_2,$
$\forall \varphi \in Y_2.$ Therefore, it suffices to show that
$\lim_{|\mathrm{Im}\lambda| \to \infty}\|\mathcal{A}_2(\lambda)\|=0$
for any $\mathrm{Re}\lambda=-\Sigma+\omega.$ Performing the change
of variables $u \mapsto \varrho=\sqrt{u^2-z^2+\eta^2}$, one sees
that
$\mathcal{A}_2^2(\lambda)=\mathcal{I}_1(\lambda)+\mathcal{I}_2(\lambda)+\mathcal{I}_3(\lambda)$
where
\begin{equation*}
\mathcal{I}_i(\lambda)=\int_0^RF_{\lambda}^i(\eta,\varrho)\psi(\varrho)\,\varrho
\,\d \varrho \qquad (i=1,2,3),
\end{equation*}
with
\begin{multline*}
F_{\lambda}^i(\eta,\varrho)=\int_{-\eta}^{\eta}
g_i(\eta,\varrho,z)e^{-(\lambda+\Sigma)\left[(2n+2)\sqrt{R^2+z^2-\eta^2}-\sqrt{\varrho^2+z^2-\eta^2}+z\right]}\theta_1(z/\eta)\times\\
\times
\dfrac{\beta_2(\sqrt{\varrho^2+z^2-\eta^2}/\varrho)}{\sqrt{\varrho^2+z^2-\eta^2}}\dfrac{\d
z}{\eta},
\end{multline*}
where
$g_1(\eta,\varrho,z)=\chi_{]0,\eta[}(\varrho)\chi_{]-\eta,-\sqrt{\eta^2-\varrho^2}[}(z),$
 $g_2(\eta,\varrho,z)=\chi_{]0,\eta[}(\varrho)\chi_{]\sqrt{\eta^2-\varrho^2},
\eta[}(z)$ and\\
$g_3(\eta,\varrho,z)=\chi_{]\eta,
R[}(\varrho)\chi_{]-\eta,\eta[}(z).$

As in the Step 3.1, one notes that, according to the
Riemann--Lebesgue Lemma \ref{riemann}, for a. e. $(\eta,\varrho) \in
(0,R) \times (0,R)$, $\lim_{|\mathrm{Im}\lambda| \to \infty}
F_{\lambda}^i(\eta,\varrho)=0$ and, since
$$\displaystyle \int_0^R \eta \,\d \eta
\int_0^R\sup_{\mathrm{Re}\lambda=-\Sigma+\omega}\left|F_{\lambda}^i(\eta,\varrho)\right|^2
\varrho\,\d \varrho < \infty,$$ the Dominated Convergence Theorem
implies
$$\lim_{|\mathrm{Im}\lambda|\to \infty}\int_0^R \eta \, \d \eta
\int_0^R\left|F_{\lambda}^i(\eta,\varrho)\right|^2 \varrho\,\d
\varrho=0,\qquad (\mathrm{Re}\lambda=-\Sigma+\omega,\:i=1,2,3).$$ We
conclude by noting that $\|\mathcal{I}_i(\lambda)\| \leq
\|F_{\lambda}^i(\cdot,\cdot)\|_{Y_2 \times Y_2}.$ This proves that
$\lim_{|\mathrm{Im}\lambda| \to
\infty}\|\mathcal{A}_2^2(\lambda)\|=0$ for any $\mathrm{Re}\lambda
>-\Sigma.$ One proceeds in the same way for
$\mathcal{A}_2^1(\lambda)$. The proof of Proposition \ref{iml2}
follows then by compiling all the above
steps as in the proof of Proposition \ref{iml}.\\

\noindent {\bf Acknowledgments.} The research of the first author
was supported by a \textit{Marie Curie Intra--European Fellowship}
within the 6th E. C. Framework Programm. The authors warmly thank
Prof. M. Mokhtar--Kharroubi for suggesting us this problem.

\bibliographystyle{plain}
{\small

\end{document}